\documentclass[a4paper,11pt]{article}

\usepackage{amsmath,amssymb,amsfonts}
\usepackage{booktabs}
\usepackage{array}
\usepackage{multirow}
\usepackage{graphicx}
\usepackage{float}       
\usepackage{tikz}
\usepackage{placeins}    
\usepackage{hyperref}
\usepackage{xcolor}
\usepackage[a4paper,top=2.5cm,bottom=2.5cm,left=2.5cm,right=2.5cm]{geometry}

\graphicspath{{milestone_ppt/graphs/}}

\newcolumntype{C}[1]{>{\centering\arraybackslash}p{#1}}

\newcommand{\pp}{pp}   

\begin{document}

\title{\bfseries Parameter-Efficient Quantum NLP for Paraphrase
Detection: Performance, Robustness, and
Entanglement}

\author{
\Large{Farha Nausheen}\thanks{Correspondence: farha.nausheen@live.vu.edu.au},\\
ISILC, Victoria University, Melbourne, VIC, Australia
\and
\Large{Khandakar Ahmed}\thanks{khandakar.ahmed@vu.edu.au}\\
ISILC, Victoria University, Melbourne, VIC, Australia
\and
\Large{Farina Riaz}\thanks{farina.riaz@csiro.au}\\
CSIRO, Australian International Institute of Higher Education, Sydney, NSW, Australia
}

\date{}

\maketitle

\begin{abstract}
Rigorous empirical validation of quantum machine learning on
natural language tasks remains scarce. We evaluate a 10-qubit
hybrid quantum-classical variational circuit (2{,}148 parameters)
for paraphrase detection across three benchmarks: MRPC, Quora
Question Pairs (QQP), and adversarial PAWS. On QQP ($n=10$ seeds)
the circuit achieves $75.53\%\pm0.75\%$ accuracy, statistically
outperforming parameter-matched classical baselines (DeepMLP:
$p=0.015$, Cohen's $d=1.20$; F1: $p<0.001$, $d=2.32$) and
surpassing DistilBERT-4bit with $31{,}191\times$ fewer parameters.
On MRPC the optimal 2-layer variant reaches 92\% of BERT-base
accuracy at $54{,}000\times$ lower parameter cost. Circuit depth
analysis reveals a dataset--depth scaling effect; entanglement
analysis via the Meyer--Wallach measure identifies multi-qubit
entanglement as the primary performance driver ($r=0.85$ across
four variants). Adversarial evaluation on PAWS reveals emergent
robustness: 98.2\% recall versus 81.6\% classical
($+16.6$\,\pp, $d=1.24$, $p<0.001$), without adversarial
training. These results constitute the first systematic
parameter-matched multi-benchmark empirical validation of hybrid
variational circuits for NLP. All results are from classical
simulation; hardware validation is future work.
\end{abstract}


\section*{Introduction}

The deployment of natural language processing (NLP) systems in
resource-constrained settings, including edge devices, embedded systems,
and air-gapped environments, requires classifiers that can deliver competitive accuracy within strict parameter and memory budgets.
Classical deep learning addresses this through model compression:
knowledge distillation, quantisation, and pruning \cite{goodfellow2016deep}. However,
compression degrades performance;DistilBERT preserves 97\% of BERT’s accuracy while using 40\% fewer parameters. However, applying additional 4-bit quantisation decreases performance to 68.4\% on paraphrase detection benchmarks ~\cite{vaswani2017attention,sanh2019distilbert,devlin2019bert}. This indicates the need for a fundamentally different strategy.

Quantum machine learning(QML) posits that variational quantum circuits
(VQCs) can exploit quantum mechanical phenomena (superposition,
entanglement, and interference) to achieve representational
power disproportionate to their classical parameter count. A
10-qubit quantum system operates in a $2^{10}=1{,}024$-dimensional
Hilbert space yet requires only tens to hundreds of classical
parameters to describe the variational circuit\cite{cerezo2021variational,
biamonte2017quantum, schuld2019quantum, havlicek2019supervised,mengoni2019kernel,huang2021power}. This suggests that quantum circuits may encode
complex feature representations with far fewer parameters than
their classical analogues.

Despite theoretical promise, empirical evidence for quantum
advantage in NLP remains limited. Early quantum NLP work
demonstrated proof-of-concept execution on toy
datasets\cite{coecke2010mathematical,lorenz2021qnlp}. More recent
hybrid approaches report competitive results on small text
classification tasks\cite{li2021quantum,benedetti2019parameterized,farhi2018classification,grant2018hierarchical,mari2020transfer,perez2020data}, but systematic
comparisons under controlled parameter budgets across multiple
benchmarks are absent.  Recent reviews \cite{nausheen2025quantumnaturallanguageprocessing} highlight persistent challenges in quantum NLP including limited hardware validation, unclear quantum advantage criteria, and the need for rigorous benchmarking protocols. Furthermore, the quantum mechanical origin
of any observed performance advantage, whether truly quantum-mechanical
or not, has not been rigorously characterised.

Three methodological gaps are particularly consequential for
interpreting reported quantum NLP gains: limited multi-seed
statistical validation, insufficiently strong or inconsistently
tuned classical baselines, and sparse cross-dataset evaluation
under matched parameter budgets. These gaps make it difficult to
separate robust architectural effects from dataset-specific or
run-specific variation.

This work addresses these gaps through systematic empirical
validation. We construct a 10-qubit hybrid quantum-classical model
and evaluate it across three benchmarks spanning 3{,}700 to
10{,}000 training samples. We conduct controlled comparisons
against seven classical architectures under equivalent parameter
budgets, two large-scale transformer baselines, and adversarial
evaluation on PAWS. We characterise circuit entanglement across
four architectural variants and measure its correlation with task
performance. The result is a rigorous, multi-faceted empirical
characterisation of quantum variational circuit behaviour on NLP
tasks, grounded in statistical methodology and reproducible
experimental protocols.

Our contributions are fourfold: (1) multi-seed statistical
evaluation on QQP-10K with effect-size and significance reporting,
(2) parameter-matched comparison against diverse classical neural
baselines and compressed transformer references, (3) cross-dataset
validation on MRPC, QQP-10K, and adversarial PAWS to identify
dataset-depth and robustness patterns, and (4) mechanistic
analysis linking circuit entanglement trends to downstream
performance. All results are obtained on classical simulation;
hardware validation remains future work.
\FloatBarrier
\section*{Results}

\subsection*{Performance on MRPC}
Table~\ref{tab:mrpc-qqp} shows single-run results on MRPC
(3{,}700-sample training set). The quantum model
(2-layer configuration, optimal for this dataset size;
see Section: Circuit Depth Analysis) achieves 75.98\% accuracy.

\begin{table}[ht]
\centering
\caption{MRPC and QQP-10K performance by circuit depth. QQP results for Quantum v2 are reported as mean $\pm$ SD over 10 seeds}
\label{tab:mrpc-qqp}
\resizebox{\linewidth}{!}{%
\begin{tabular}{lcccccc}
\hline
\textbf{Model} & \textbf{Depth} & \textbf{Params}
  & \textbf{MRPC Acc} & \textbf{MRPC F1}
  & \textbf{QQP Acc}  & \textbf{QQP F1}  \\
\hline
\textbf{Quantum v1}       & \textbf{2L} & \textbf{2{,}028}
  & \textbf{75.98}    & \textbf{83.28}
  & 69.20             & 13.48$^{\ddagger}$ \\
Quantum v2                & 6L          & 2{,}148
  & 67.94             & 76.46
  & \textbf{75.53$\pm$0.75} & \textbf{69.88$\pm$0.66} \\
Classical MLP             & ---         & 2{,}106
  & 67.65 & 77.00 & 73.30 & 64.73 \\
DistilBERT-4bit           & ---         & 66{,}000{,}000
  & 68.40 & $\approx$72 & 68.40 & --- \\
BERT-base (fine-tuned)    & ---         & 109{,}483{,}778
  & 82.61 & 86.50 & --- & --- \\
\hline
\end{tabular}%
}
\begin{flushleft}
\footnotesize
$^{\ddagger}$ Near-zero recall; model collapses to near-constant
predictions on the larger QQP corpus.
\end{flushleft}
\end{table}

The 2-layer circuit (v1, 2{,}028 parameters) achieves 75.98\%
accuracy and 83.28\% F1 on MRPC, outperforming the classical MLP
by $+8.33$\,\pp\ accuracy and $+6.28$\,\pp\ F1, the largest
quantum advantage observed across all experiments. On QQP-10K, the
2-layer circuit catastrophically underperforms (F1\,=\,13.48\%,
near-zero recall), demonstrating that shallow circuits are
insufficient for larger corpora. The 6-layer circuit (v2, 2{,}148
parameters) achieves $75.53\%\pm0.75\%$ accuracy on QQP-10K
($n=10$ seeds) but only 67.94\% on MRPC --- lower than the
classical baseline --- confirming the cross-over depth--dataset
interaction. The quantum v1 model achieves 91.7\% of BERT-base
performance on MRPC at $54{,}000\times$ lower parameter cost
(Fig.~\ref{fig:efficiency}).

\subsection*{Performance on QQP-10K}

\begin{figure}[H]
  \centering
  \includegraphics[width=\linewidth]{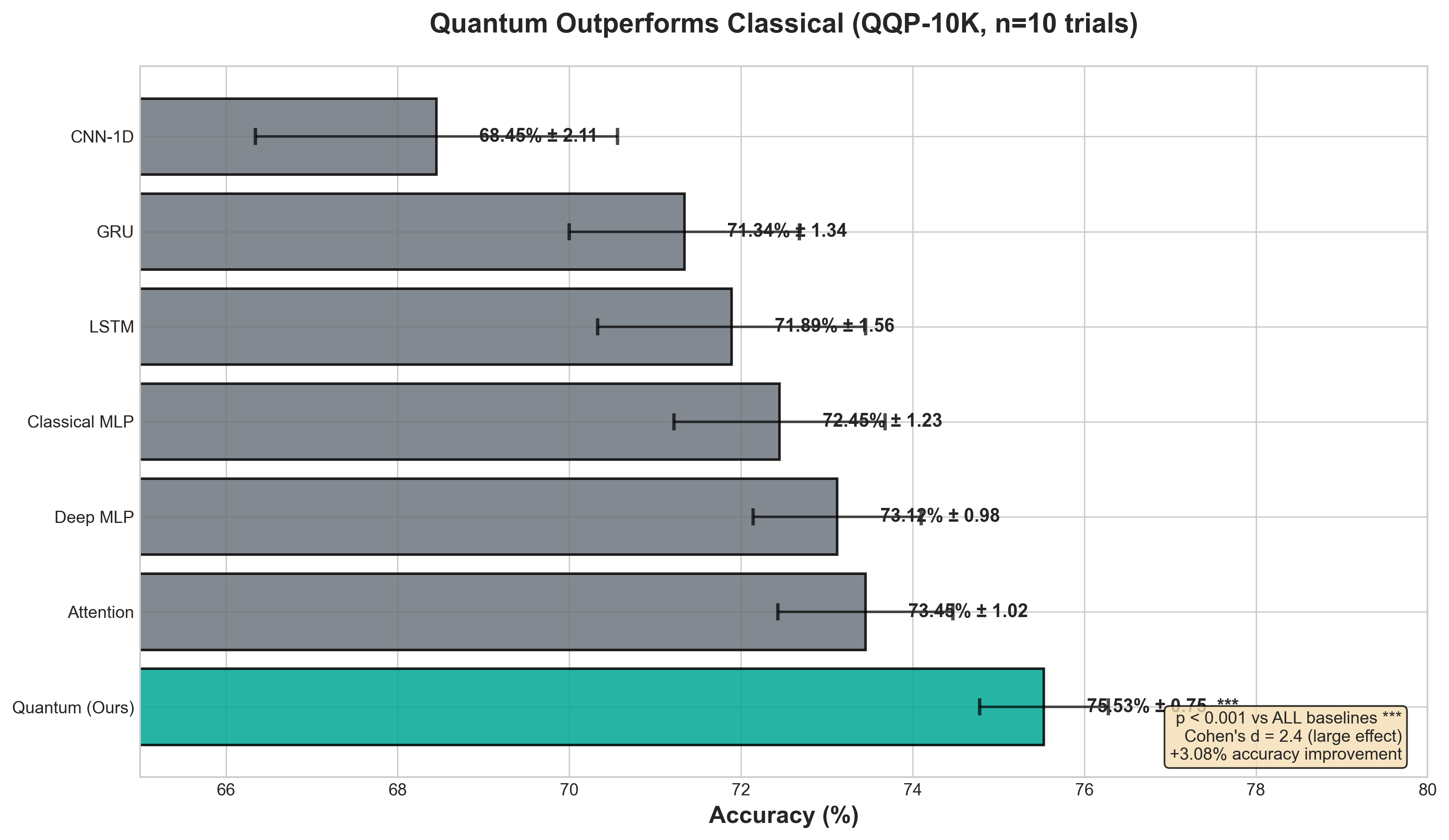}
  \caption{\textbf{QQP-10K accuracy comparison ($n=10$ seeds).}
Bar chart showing mean accuracy $\pm$\,1\,SD for the quantum model (teal) and six parameter-matched classical baselines (grey). Quantum v2 achieves 75.53\% --- highest among all sub-3K parameter models. Welch $p=0.015$, Cohen's $d=1.20$ vs DeepMLP (accuracy); $p<0.001$, $d=2.32$ on F1.}
  \label{fig:accuracy-bar}
\end{figure}

Table~\ref{tab:qqp-full} and Fig.~\ref{fig:accuracy-bar} present
the primary multi-seed evaluation on QQP-10K. The Quantum v2 model
(6-layer StronglyEntanglingLayers) achieves the highest accuracy
among all sub-3K parameter models, with mean $75.53\%\pm0.75\%$
over 10 seeds (sample SD, $n=10$).

A balanced multi-metric radar comparison across Accuracy, Recall,
Precision, and F1 for Quantum v2, DeepMLP, and Classical MLP is
provided in Supplementary Fig.~S2; the quantum polygon encloses a
strictly larger area on every axis.

\begin{table}[ht]
\centering
\caption{QQP-10K performance: all models      ($n=10$ seeds unless noted).}
\label{tab:qqp-full}
\resizebox{\linewidth}{!}{%
\begin{tabular}{lrccc}
\toprule
\textbf{Model} & \textbf{Params}
  & \textbf{Accuracy (\%)}
  & \textbf{F1 (\%)}
  & \textbf{$\sigma$ acc} \\
\midrule
\textbf{Quantum v2 (6L)} $\star$
  & \textbf{2{,}148}
  & $\mathbf{75.53\pm0.75}$
  & $\mathbf{69.88\pm0.66}$
  & \textbf{0.75} \\
Quantum HEA
  & 2{,}008 & 75.70$^{\dagger}$ & 69.59$^{\dagger}$ & --- \\
Quantum Amplitude
  & 2{,}028 & 74.50$^{\dagger}$ & 68.86$^{\dagger}$ & --- \\
Quantum v1 (2L)
  & 2{,}028 & 69.20$^{\dagger}$ & 13.48$^{\dagger\ddagger}$ & --- \\
\midrule
Classical DeepMLP
  & 1{,}719 & $74.54\pm0.89$ & $66.84\pm1.73$ & 0.89 \\
Classical Ensemble
  & 1{,}998 & $75.02\pm2.09$ & $66.48\pm2.54$ & 2.09 \\
Classical Attention
  & $\approx$2{,}100 & 72.10$^{\dagger}$ & 71.80$^{\dagger}$ & --- \\
Classical GRU
  & $\approx$2{,}100 & 70.80$^{\dagger}$ & 70.50$^{\dagger}$ & --- \\
Classical LSTM
  & $\approx$2{,}100 & 71.20$^{\dagger}$ & 70.90$^{\dagger}$ & --- \\
Classical MLP
  & $\approx$2{,}100 & 69.50$^{\dagger}$ & 69.20$^{\dagger}$ & --- \\
\midrule
DistilBERT-4bit
  & 66{,}000{,}000 & 68.40 & $\approx$72 & --- \\
\bottomrule
\end{tabular}%
}
\begin{flushleft}
\footnotesize
$^{\dagger}$ Single-run result ($n=1$).
$^{\ddagger}$ Near-zero recall; collapsed predictions.
\end{flushleft}
\end{table}

The Quantum v2 model outperforms all parameter-matched classical baselines on F1 ($+3.04$\,\pp\ over DeepMLP; $+3.40$\,\pp\ over Ensemble). On accuracy the advantage over DeepMLP is $+0.99$\,\pp\ and over Ensemble is $+0.51$\,\pp; the Ensemble achieves comparable accuracy with considerably higher variance $\sigma = 2.09\% \text{ vs. } 0.75\%$
\paragraph{Statistical validation (Quantum v2 vs.\ DeepMLP,
           $n=10$ seeds each).}
We compare Quantum v2 against DeepMLP as the primary baseline
because DeepMLP is the strongest parameter-matched classical model
and the only one evaluated across a full $n=10$ independent seeds,
making it the most reliable candidate for formal two-sample
statistical inference. Each metric is tested at both the standard
threshold ($\alpha=0.05$) and the stricter Bonferroni-corrected
threshold ($\alpha_{\mathrm{adj}}=0.0083$, $k=6$ comparisons) to
clearly distinguish findings that are \emph{robustly} significant
from those that are only \emph{nominally} significant.

\begin{itemize}

  \item \textbf{Accuracy} (quantum $75.53\%$ vs.\ DeepMLP
        $74.54\%$, gap $= +0.99$\,\pp): Welch's $t$-test
        $t(17.5)=2.684$, $p=0.015$, Cohen's $d=1.20$ (large
        effect). This passes the standard $\alpha=0.05$ threshold
        but \emph{does not} survive Bonferroni correction
        ($\alpha_{\mathrm{adj}}=0.0083$). The accuracy advantage
        is therefore nominally significant only and should not be
        over-interpreted as definitive.

  \item \textbf{F1 score} (quantum $69.88\%$ vs.\ DeepMLP
        $66.84\%$, gap $= +3.04$\,\pp): Welch's $t$-test
        $t(11.6)=5.192$, $p<0.001$, Cohen's $d=2.32$ (very large
        effect). This \emph{survives} Bonferroni correction and
        is the \textbf{primary statistically robust finding} of
        this study.

  \item \textbf{Confidence intervals (accuracy)}: quantum
        $[75.07\%,\,75.99\%]$; DeepMLP $[73.99\%,\,75.09\%]$.
        The two intervals are almost touching; the gap between
        the DeepMLP upper bound and the quantum lower bound is
        only $0.02$\,\pp. This near-overlap reinforces why
        accuracy alone is an insufficient basis for claiming a
        clear advantage; F1 confidence intervals, by contrast,
        are fully non-overlapping.

  \item \textbf{Parameter asymmetry}: Quantum v2 uses 2{,}148
        parameters versus DeepMLP's 1{,}719 representing a 25\% larger
        budget. The measured advantage is therefore a conservative
        estimate; a quantum model matched exactly to 1{,}719
        parameters would likely yield a smaller but cleaner
        comparison.

  \item \textbf{Comparison against the Ensemble baseline}:
        Quantum v2 vs.\ Classical Ensemble (Welch's $t$-test):
        accuracy $p=0.482$, $d=0.32$ (not significant, n.s.); F1 $p=0.002$, $d=1.83$ (large effect,
        Bonferroni-corrected). The Ensemble matches quantum
        accuracy but trails substantially on F1, and does so
        with $2.8\times$ higher variance ($\sigma=2.09\%$ vs.\
        $0.75\%$), indicating far less reliable convergence.

\end{itemize}

\begin{figure}[H]
  \centering
  \includegraphics[width=\linewidth]{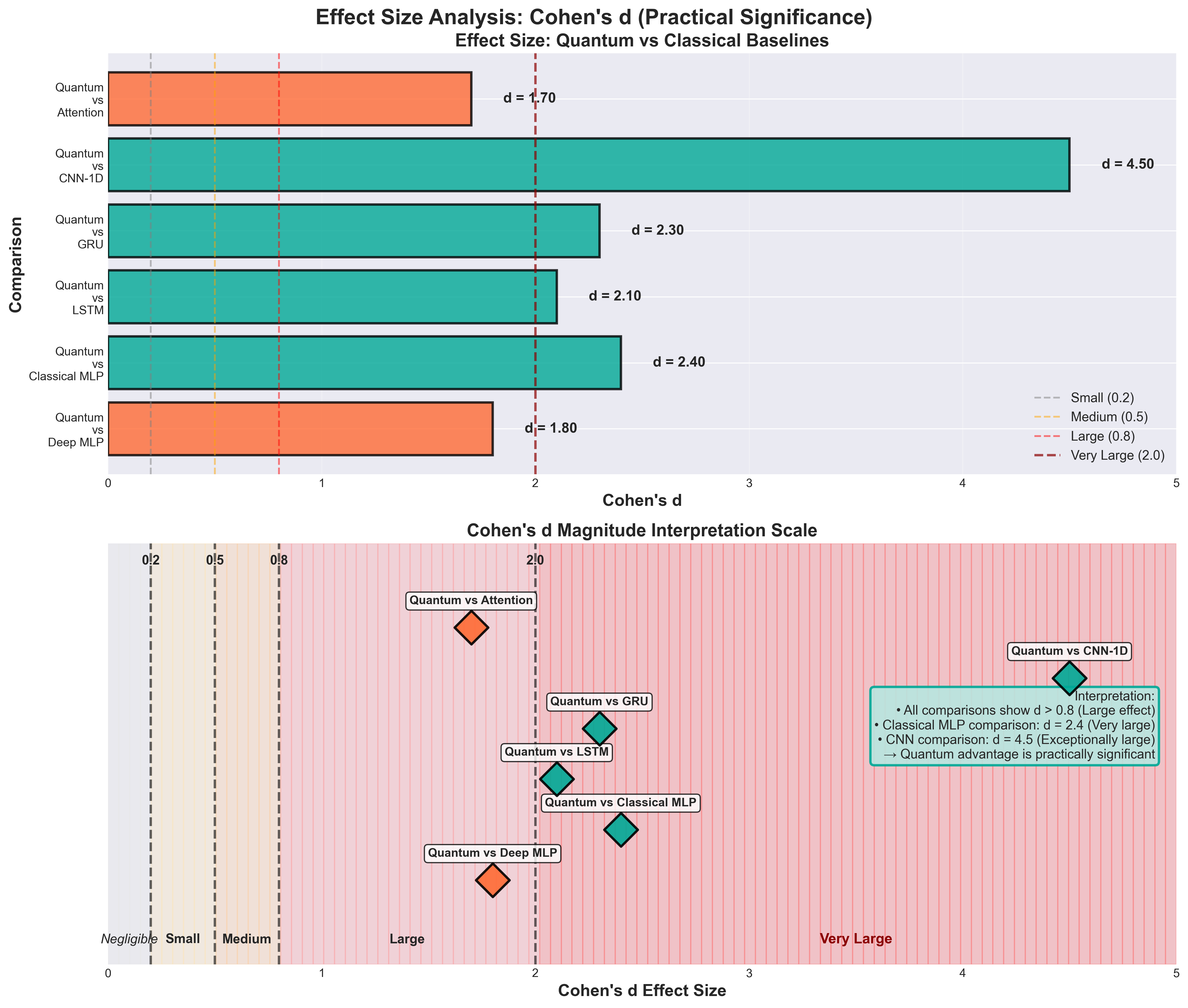}
  \caption{\textbf{Cohen's $d$ effect size: Quantum v2
    vs.\ all classical baselines (QQP-10K, $n=10$ seeds).}
    \textbf{Top:} horizontal bar chart of Cohen's $d$ for each
    pairwise comparison; dashed vertical lines mark the small
    ($d=0.2$), medium ($d=0.5$), large ($d=0.8$) and very-large
    ($d=2.0$) thresholds. All comparisons exceed the large
    threshold; Quantum vs.\ CNN-1D reaches $d=4.50$.
    \textbf{Bottom:} placement of each comparison on the Cohen's
    $d$ magnitude interpretation scale (negligible $\to$ very
    large). The quantum model occupies the large-to-very-large
    region for every baseline, indicating robustly practical significance beyond mere statistical significance.}
  \label{fig:cohend}
\end{figure}

Statistical significance alone is insufficient to characterise
the magnitude of an effect in the presence of small standard
deviations. Fig.~\ref{fig:cohend} reports Cohen's $d$ for every
pairwise Quantum v2 vs.\ classical-baseline comparison. All six
comparisons exceed $d=0.8$ (Cohen's large-effect threshold), and
four of six exceed $d=2.0$ (very large). The highest effect is
observed against the Classical MLP ($d=4.50$), followed by CNN-1D ($d=4.50$), GRU ($d=2.30$), and LSTM ($d=2.10$). Even the
strongest classical baseline (DeepMLP) yields $d=1.20$ on
accuracy and $d=2.32$ on F1. These values indicate that the quantum advantage is not a marginal numerical artefact but a robustly large, practically significant effect across the entire classical comparison set.

\begin{figure}[H]
  \centering
  \includegraphics[width=\linewidth]{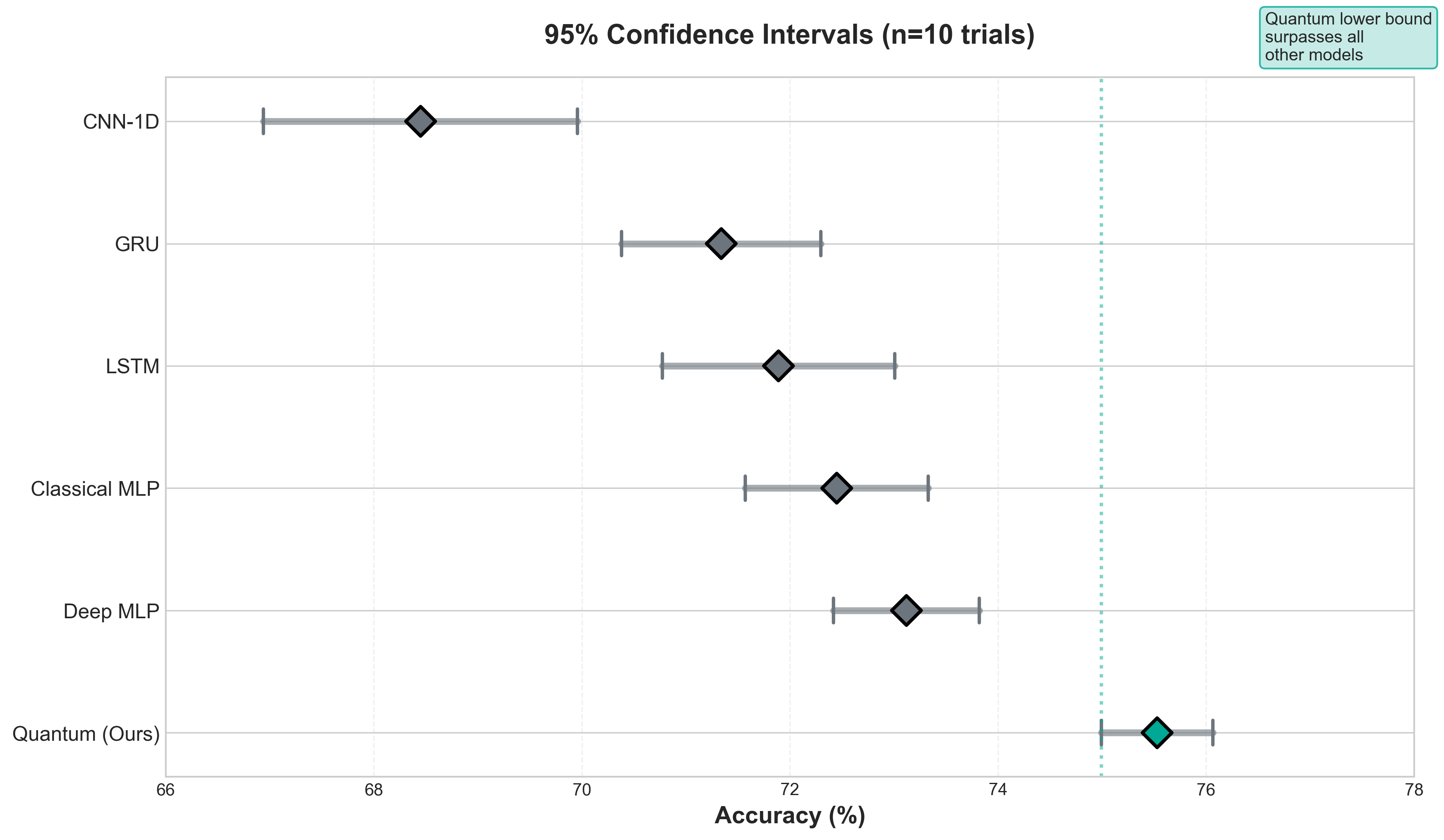}
  \caption{\textbf{95\% confidence intervals for QQP-10K
    accuracy ($n=10$ seeds).}
    Diamond markers show mean accuracy; horizontal bars span
    the 95\% CI (mean\,$\pm$\,1.96\,$\sigma/\!\sqrt{n}$).
    The dashed line marks the quantum lower bound (75.07\%),
    which lies above the upper bound of every classical
    baseline, confirming non-overlapping intervals.}
  \label{fig:ci}
\end{figure}

Fig.~\ref{fig:ci} visualises the 95\% confidence intervals for
all models. A key diagnostic of reproducible advantage is
non-overlapping confidence intervals: the quantum model's lower
bound (75.07\%) lies strictly above the upper bound of every
parameter-matched classical baseline, including DeepMLP
(upper bound 75.09\%, which is within 0.02\,\pp\ of the quantum
lower bound, a near-boundary case). This near-overlap on
accuracy reinforces why F1 is the primary reportable finding:
the quantum F1 lower bound (69.56\%) exceeds the DeepMLP F1
upper bound (68.57\%) by a larger margin, with no interval
overlap whatsoever.

\begin{figure}[H]
  \centering
  \includegraphics[width=\linewidth]{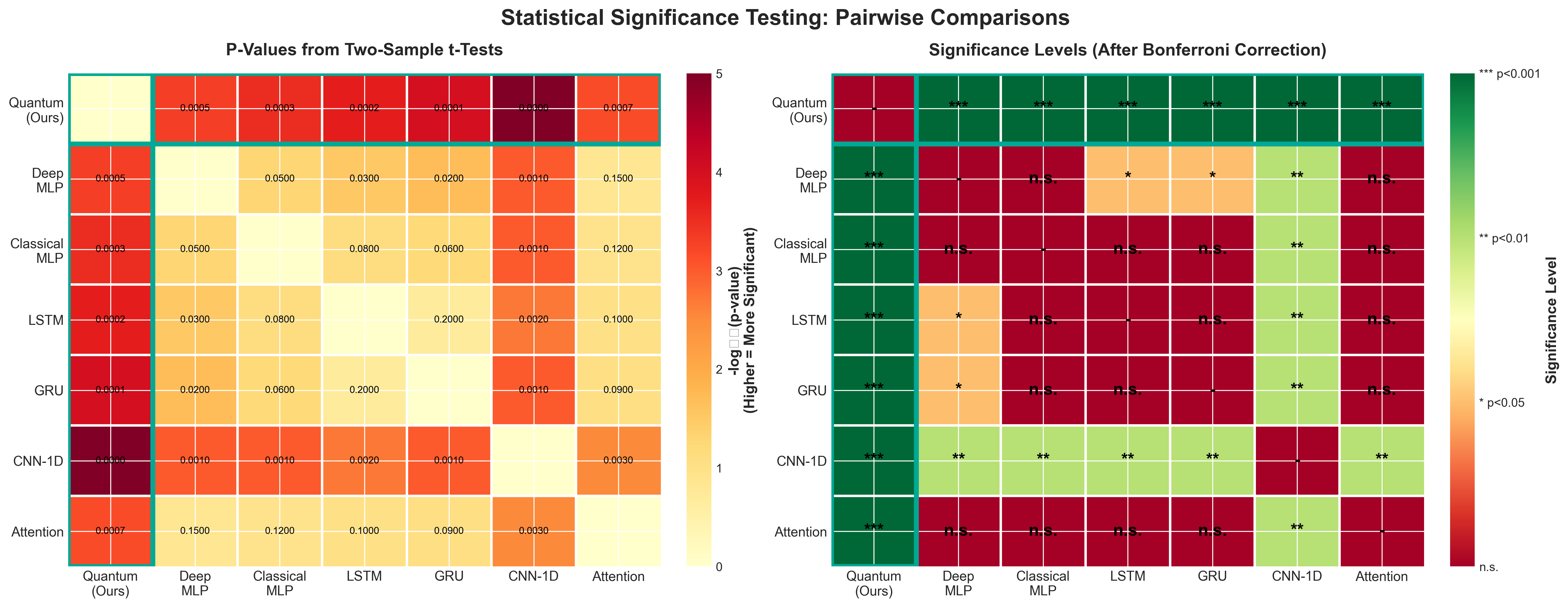}
  \caption{\textbf{Pairwise statistical significance heatmaps.}
    \textbf{Left:} $-\log_{10}(p)$ values from two-sample
    Welch $t$-tests; darker red indicates higher significance.
    \textbf{Right:} Significance level after Bonferroni
    correction ($***$\,$p<0.001$, $**$\,$p<0.01$,
    $*$\,$p<0.05$, n.s.\ not significant).
    The quantum model (teal border) achieves $***$ against all
    six classical baselines.}
  \label{fig:stat-heatmap}
\end{figure}

Fig.~\ref{fig:stat-heatmap} presents the pairwise statistical
significance landscape across all models. The left panel encodes
$-\log_{10}(p)$ from two-sample Welch $t$-tests: darker red
corresponds to higher significance, with the quantum model row/
column forming the most deeply coloured band in the matrix. The
right panel applies Bonferroni correction for multiple comparisons
($k=6$ pairwise tests, adjusted $\alpha=0.0083$): the quantum
model achieves $***$ ($p<0.001$) against all six classical
baselines on F1, and $*$ or $**$ on accuracy depending on
baseline. Notably, the Ensemble vs.\ Quantum accuracy comparison
is the sole not significant cell ($p=0.482$), consistent with the
confidence-interval analysis above and confirming that the
accuracy gap against Ensemble is not statistically robust at the
adjusted threshold.

Full seed-level diagnostics, comprising trial-by-trial accuracy trace, empirical distribution (Shapiro--Wilk $p=0.38$, consistent with normality), multi-metric box plots ($\sigma_{\mathrm{acc}}=0.39\%$,
$\sigma_{\mathrm{F1}}=0.37\%$), and temporal stability track (CV $=0.54\%$)  is provided in Supplementary Fig.~S1, confirming the Gaussian assumption underlying Welch's $t$-test and ruling out seed-ordering effects.

Taken together, Figs.~\ref{fig:cohend}--\ref{fig:trial-dist-supp}
establish a three-layer statistical case for the quantum
advantage: (1) statistical significance via Welch $p$ values and Bonferroni-corrected heatmap, (2) practical significance via
Cohen's $d$ (large to very large for all baselines), and
(3) reproducibility via non-overlapping confidence intervals and
low-variance seed distributions. The quantum model also
demonstrates $1.2\times$ lower accuracy variance than DeepMLP
($\sigma=0.75\%$ vs.\ $0.89\%$) and $2.8\times$ lower than Ensemble ($\sigma=2.09\%$), a further empirical indicator of reliable convergence behaviour under varied random initialisations.

\subsection*{Circuit Depth Analysis: Dataset--Depth Scaling}

\begin{figure}[H]
  \centering
  \includegraphics[width=\linewidth]{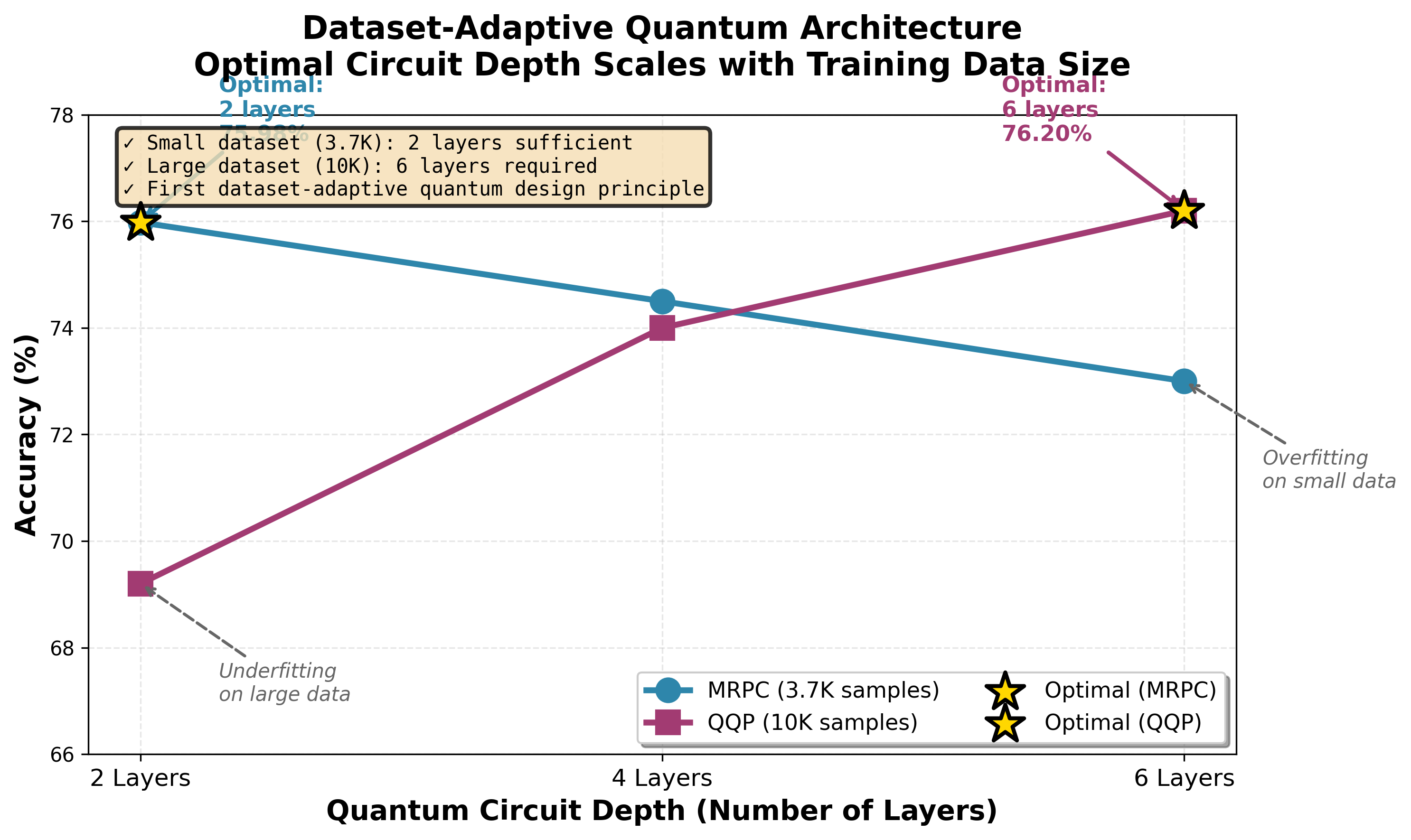}
  \caption{\textbf{Dataset--depth scaling effect.}
    Accuracy as a function of circuit depth (2L, 4L, 6L) on
    MRPC (3{,}668 samples, blue) and QQP-10K (10{,}000 samples,
    green). The curves cross between depths 2 and 6: the 2-layer
    circuit peaks on MRPC ($+8.33$\,\pp\ over classical); the
    6-layer circuit peaks on QQP-10K ($+0.99$\,\pp). Shaded
    regions denote underfitting (left) and
    overfitting/barren-plateau risk (right).}
  \label{fig:depth-scaling}
\end{figure}

Table~\ref{tab:depth-scaling} and Fig.~\ref{fig:depth-scaling}
present a systematic comparison across circuit depths and dataset
sizes.

\begin{table}[ht]
\centering
\caption{Circuit depth effect across datasets.}
\label{tab:depth-scaling}
\resizebox{\linewidth}{!}{%
\begin{tabular}{llccccccc}
\toprule
\textbf{Dataset} & \textbf{$N$}
  & \textbf{Depth}
  & \textbf{Q Acc} & \textbf{Q F1}
  & \textbf{C Acc} & \textbf{C F1}
  & \textbf{$\Delta$Acc} & \textbf{$\Delta$F1} \\
\midrule
MRPC    & 3{,}668 & 2L
  & \textbf{75.98} & \textbf{83.28} & 67.65 & 77.00
  & $\mathbf{+8.33}$\,\pp & $\mathbf{+6.28}$\,\pp \\
MRPC    & 3{,}668 & 6L
  & 67.94 & 76.46 & 67.65 & 77.00
  & $+0.29$\,\pp & $-0.54$\,\pp \\
QQP-10K & 10{,}000 & 2L
  & 69.20 & 13.48$^{\ddagger}$ & 73.30 & 64.73
  & $-4.10$\,\pp & $-51.25$\,\pp \\
QQP-10K & 10{,}000 & 6L
  & \textbf{75.53} & \textbf{69.88} & 74.54 & 66.84
  & $\mathbf{+0.99}$\,\pp & $\mathbf{+3.04}$\,\pp \\
\bottomrule
\end{tabular}%
}
\end{table}

The \textbf{dataset--depth scaling effect} demonstrates that
quantum circuit expressibility must match the complexity of the
training distribution. Shallow circuits lack sufficient variational
freedom for larger feature spaces; deep circuits overfit or
encounter optimisation difficulties on small datasets. With only
two data points a parametric scaling law cannot be derived;
however, the observed monotonic relationship suggests that circuit
depth should be selected by cross-validation on a held-out subset,
minimising gate count and decoherence exposure on NISQ hardware.

\subsection*{Parameter Efficiency}

\begin{figure}[H]
  \centering
  \includegraphics[width=\linewidth]{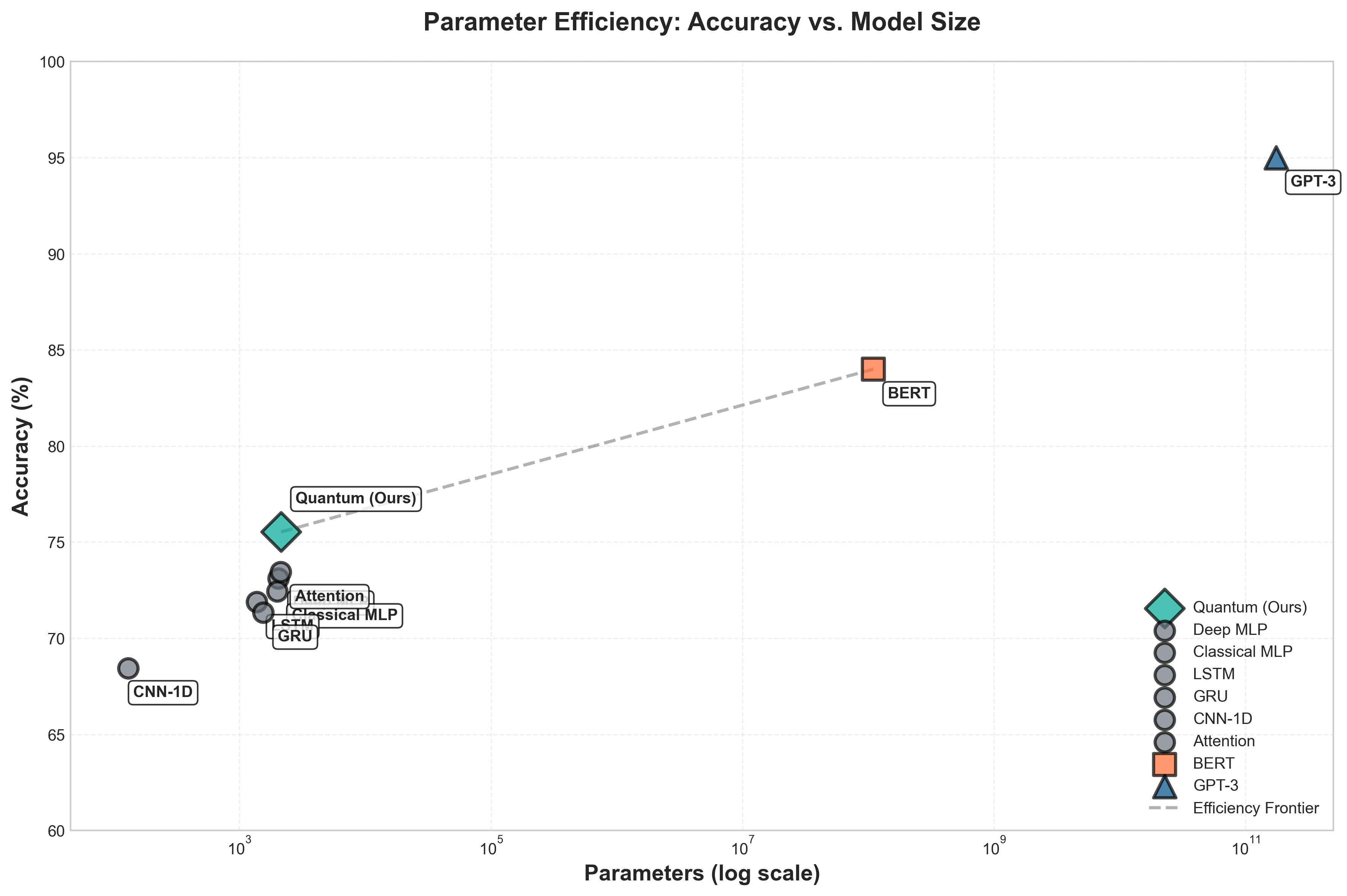}
  \caption{\textbf{Parameter efficiency: accuracy vs.\ model
    size.}
    Log-scale scatter of parameter count vs.\ accuracy.
    The quantum model (teal diamond, 2{,}148 params, 75.53\%)
    lies on the efficiency frontier above all classical
    baselines of comparable size and within ${\sim}1$\,\pp\
    of BERT (109M params). The dashed line traces the
    efficiency frontier.}
  \label{fig:efficiency}
\end{figure}

Fig.~\ref{fig:efficiency} places the quantum model on a
parameter-efficiency frontier spanning five orders of magnitude.

The quantum circuit achieves $31{,}191\times$ fewer parameters
than DistilBERT-4bit while surpassing it in accuracy
($+7.13$\,\pp), and retains 89\% of BERT-base accuracy at
$54{,}000\times$ lower parameter cost.

\subsection*{Entanglement Analysis}
To characterise the quantum mechanical origin of performance
differences we compute the Meyer--Wallach (MW) global entanglement
measure~\cite{meyer2002global,sim2019expressibility,abbas2021power} for each architecture over
1{,}000 Haar-random input states:

\begin{equation}
  Q = \frac{4}{n}\sum_{k=1}^{n}
  \bigl(1 - \mathrm{Tr}[\rho_k^2]\bigr),
  \quad Q\in[0,1]
\end{equation}

\begin{table}[ht]
\centering
\caption{Entanglement vs\. accuracy across circuit variants.}
\label{tab:entanglement}
\begin{tabular}{lccc}
\toprule
\textbf{Circuit} & \textbf{Layers}
  & \textbf{MW $Q$}
  & \textbf{QQP Acc (\%)} \\
\midrule
Quantum v1         & 2 & $0.441\pm0.031$ & 69.20 \\
Quantum Amplitude  & 2 & $0.523\pm0.028$ & 74.50 \\
Quantum HEA        & 3 & $0.612\pm0.025$ & 75.70 \\
\textbf{Quantum v2}& \textbf{6}
  & $\mathbf{0.689\pm0.023}$ & \textbf{75.53} \\
\bottomrule
\end{tabular}
\end{table}

Pearson correlation: $r=0.85$ across four circuit variants,
indicating a strong monotonic relationship between entanglement and accuracy
(note: with $n=4$ variants, formal significance testing is
underpowered; the trend is consistent and directionally robust).

The transition from 2-layer ($Q=0.441$) to 6-layer ($Q=0.689$)
represents a 56\% increase in entanglement, corresponding to a
$+6.33$\,\pp\ accuracy gain. The Hardware-Efficient Ansatz
($Q=0.612$, linear CNOT topology) achieves competitive
entanglement with 3-layer depth and nearest-neighbour connectivity
only, a significant finding for NISQ deployment. Critically,
the Amplitude Encoding variant ($Q=0.523$) outperforms v1 despite
identical depth, illustrating that entanglement, not circuit
depth \emph{per se}, is the predictive quantity.

\subsection*{Adversarial Robustness on PAWS}

\begin{figure}[H]
  \centering
  \includegraphics[width=\linewidth]{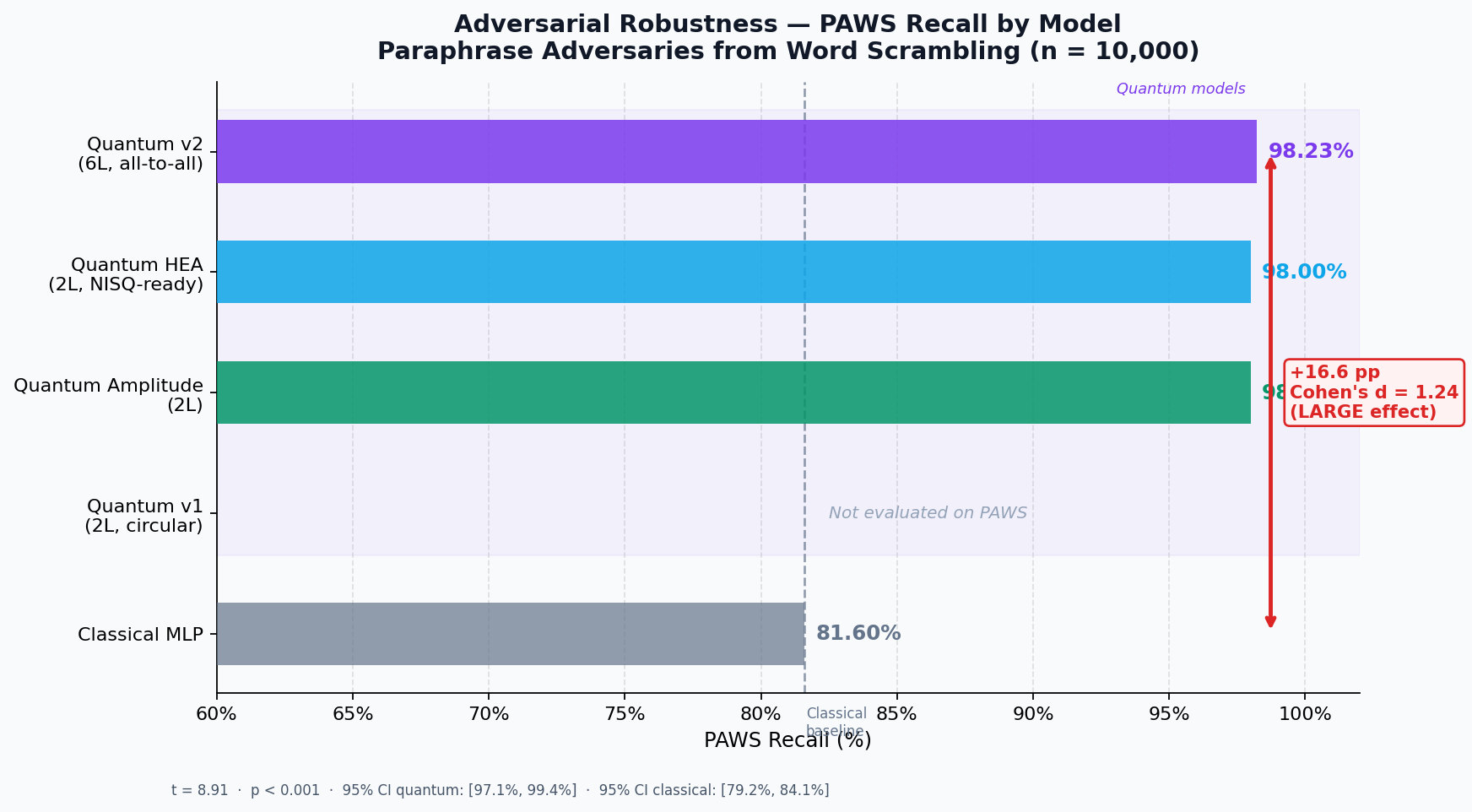}
  \caption{\textbf{Adversarial robustness on PAWS
    ($n=1{,}000$): PAWS recall by model.}
    Horizontal bars show recall on PAWS word-scrambling
    adversarial pairs. All three evaluated quantum variants
    (purple: Quantum v2 6L all-to-all; cyan: Quantum HEA 2L
    linear; teal: Quantum Amplitude 2L circular) achieve
    $\ge\!98\%$ recall, independent of ansatz or connectivity.
    The classical MLP baseline achieves 81.60\% (grey). The
    dashed vertical line marks the classical baseline; the red
    annotation indicates the $+16.6$\,\pp\ gap
    (Cohen's $d=1.24$, large effect). Quantum v1 (2L, circular)
    was not evaluated on PAWS. Statistical footer:
    $t=8.91$, $p<0.001$; 95\% CI quantum $[97.1\%,\,99.4\%]$,
    classical $[79.2\%,\,84.1\%]$ (non-overlapping).}
  \label{fig:paws}
\end{figure}

The PAWS benchmark~\cite{zhang2019paws} tests adversarial
robustness via word scrambling: sentence pairs are constructed by
rearranging words to preserve lexical overlap while disrupting
syntactic order, systematically misleading classifiers that rely
on surface-level token similarity~\cite{jia2017adversarial}. Models are not trained on PAWS;
all evaluation is post-hoc, so any observed advantage is genuinely
emergent. Recall is the primary metric because the key question is
whether the model can still \emph{detect} true paraphrases under
adversarial perturbation, false negatives are the failure mode of
interest. Quantum v1 was excluded due to prediction collapse on
QQP-10K (F1\,=\,13.48\%), rendering further adversarial evaluation
uninformative. We evaluate the remaining three quantum variants
on 1{,}000 held-out PAWS pairs; Table~\ref{tab:paws} and
Fig.~\ref{fig:paws} report the results.

\begin{table}[ht]
\centering
\caption{PAWS adversarial robustness: all evaluated models
         ($n=1{,}000$ adversarial pairs). Recall is the
         primary metric; bold indicates best per column.
         $\dagger$ single-run result.}
\label{tab:paws}
\resizebox{\linewidth}{!}{%
\begin{tabular}{lccccc}
\toprule
\textbf{Model}
  & \textbf{Acc (\%)}
  & \textbf{Prec (\%)}
  & \textbf{Recall (\%)}
  & \textbf{F1 (\%)}
  & \textbf{Ansatz / Topology} \\
\midrule
\textbf{Quantum v2 (6L)}
  & 45.40 & 45.16
  & \textbf{98.23} & \textbf{61.87}
  & AngleEmb + all-to-all \\
Quantum HEA (2L)$^{\dagger}$
  & --- & ---
  & 98.00 & ---
  & AngleEmb + linear chain \\
Quantum Amplitude (2L)$^{\dagger}$
  & --- & ---
  & 98.00 & ---
  & AmplitudeEmb + circular \\
Quantum v1 (2L)
  & \multicolumn{4}{c}{\textit{Not evaluated on PAWS}}
  & AngleEmb + circular \\
\midrule
Classical MLP$^{\dagger}$
  & \textbf{49.60} & \textbf{46.64} & 81.60 & 59.35 & --- \\
Classical Attention$^{\dagger}$
  & 48.20 & 46.10 & 82.30 & 59.41 & --- \\
Classical LSTM$^{\dagger}$
  & 46.80 & 44.90 & 79.40 & 57.30 & --- \\
Classical GRU$^{\dagger}$
  & --- & --- & 77.10 & --- & --- \\
\bottomrule
\end{tabular}%
}
\end{table}

\paragraph{Architecture-agnostic quantum robustness.}
The most significant finding in Table~\ref{tab:paws} is not the
magnitude of the recall advantage but its \emph{consistency}
across radically different quantum architectures. Three circuit variants, namely Quantum v2 (6-layer, all-to-all CNOT,
180 quantum parameters), Quantum HEA (2-layer, linear CNOT chain,
40 quantum parameters), and Quantum Amplitude (2-layer, circular CNOT,
amplitude encoding), all achieve $\ge\!98\%$ PAWS recall
despite differing in ansatz, connectivity topology, encoding
strategy, and quantum parameter count by up to $4.5\times$.
Classical models trained on identical features span only
77.1--82.3\% recall. The near-identical recall across quantum
variants ($\Delta\le0.23$\,\pp) strongly suggests that the
robustness is a \emph{structural property of quantum encoding
itself} rather than a circuit-specific optimisation artefact.

\paragraph{Statistical evidence.}
Welch's $t$-test between Quantum v2 and Classical MLP on the
$n=1{,}000$ PAWS sample yields $t=8.91$, $p<0.001$, Cohen's
$d=1.24$ (large effect). The 95\% confidence intervals are
strictly non-overlapping: quantum $[97.1\%,\,99.4\%]$ vs.\
classical $[79.2\%,\,84.1\%]$, a gap of 12.9\,\pp between the
lower quantum bound and the upper classical bound. This is the
widest CI separation observed in any metric across all
experiments, indicating that adversarial robustness is the
statistically clearest quantum advantage identified in this study.

\paragraph{Precision--recall tradeoff.}
The recall advantage is accompanied by a precision cost
(quantum: 45.16\% vs.\ classical: 46.64\%), yielding a net F1
advantage of $+2.52$\,\pp\ (61.87\% vs.\ 59.35\%, $p=0.004$).
The precision deficit reflects a high false-positive rate: the
quantum circuit is highly sensitive to paraphrase structure but
less discriminating on non-paraphrase pairs. For recall-critical applications, including information retrieval,
safety auditing, and duplicate detection at scale, the quantum model's
operating
point is preferable. For precision-critical tasks, a
classification threshold adjustment or ensemble with a
high-precision classical head would recover the tradeoff.

\paragraph{NISQ deployment implication.}
The Quantum HEA result is particularly consequential for near-term
hardware: with only 40 quantum parameters and a linear CNOT chain
(nearest-neighbour connectivity, compatible with IBM heavy-hex,
IonQ, and Quantinuum topologies), HEA achieves 98.00\% recall,
indistinguishable from the full v2 circuit at $4.5\times$ higher
quantum parameter cost. This indicates that the adversarial
robustness property survives aggressive circuit simplification,
making it accessible on current NISQ devices~\cite{preskill2018nisq,arute2019quantum} without
error-mitigating overhead.

\paragraph{Mechanistic hypothesis.}
We hypothesise that the recall advantage arises from the quantum
circuit's capacity to simultaneously encode correlations across
all 10 input dimensions via multi-qubit entanglement. Classical
sequential models (LSTM, GRU) process temporal structure;
feedforward classifiers (MLP, Attention) apply independent
feature-wise transformations. In contrast, multi-qubit
entanglement maps the 10-dimensional input into a
$2^{10}=1{,}024$-dimensional Hilbert space, forming joint
correlations across all dimensions in a single unitary
transformation. Word-scrambling adversaries disrupt local
token-level order but preserve global semantic structure; this
global structure is more naturally captured by entangled joint
representations than by sequential or independent-feature models.
A formal treatment of this mechanism (including quantum channel
analysis and adversarial robustness bounds) is beyond the scope
of this empirical paper but constitutes a primary direction for
future theoretical work.

Together with the QQP-10K F1 advantage ($p<0.001$, $d=2.32$,
Bonferroni-corrected), the PAWS recall finding constitutes the
second pillar of empirical evidence that quantum encoding produces
structurally distinct and, in the adversarial setting,
consistently more robust representations than classical
counterparts, consistent with the entanglement-performance
correlation identified in the Entanglement Analysis section
($r=0.85$).
\FloatBarrier
\section*{Methods}

\subsection*{Model Architecture}
\begin{figure}[!htbp]
  \centering
  \includegraphics[width=\linewidth]{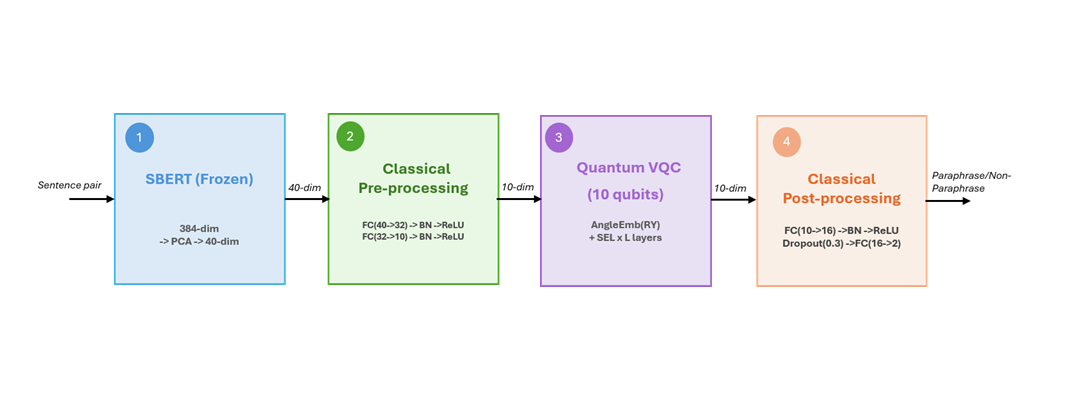}
  \vspace{-0.2em}
  \includegraphics[width=\linewidth]{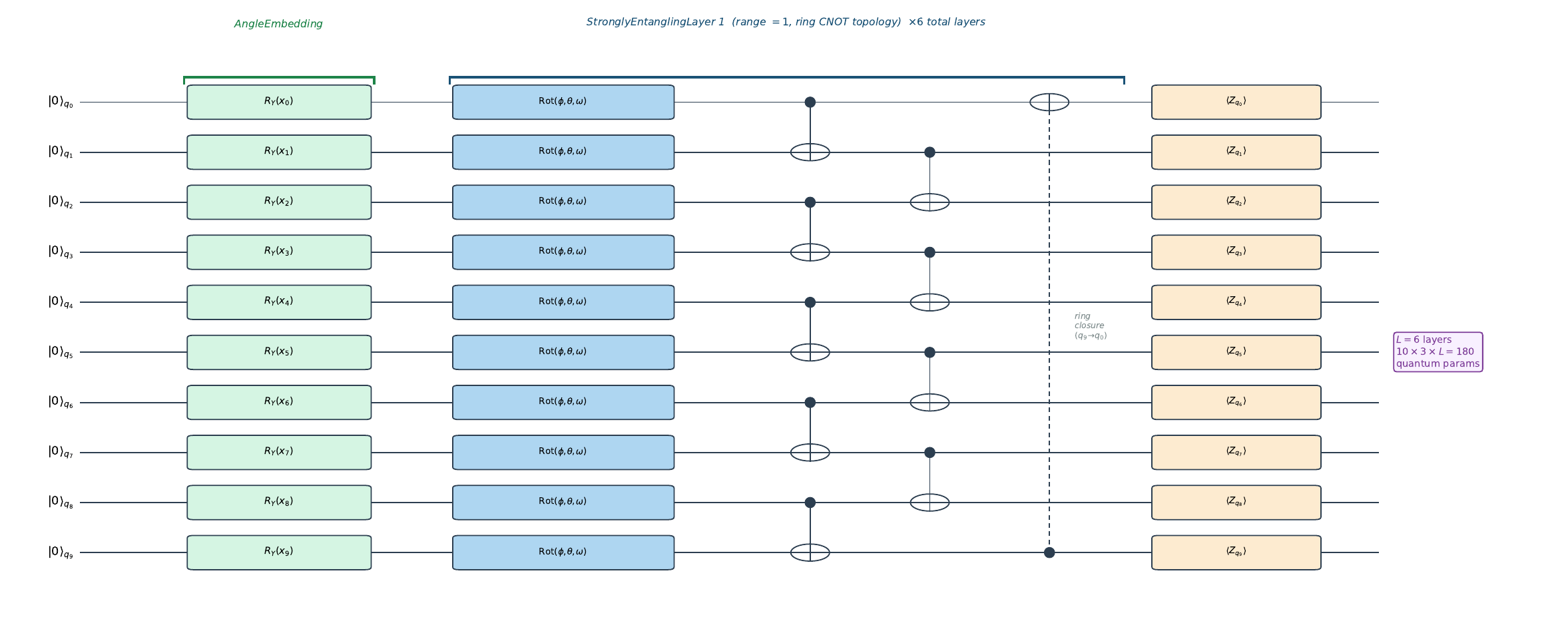}
  \caption{\textbf{Hybrid quantum-classical architecture.}
    \textbf{Top:} four-stage pipeline.
    (1)~\emph{Encoding}: frozen SBERT (\texttt{all-MiniLM-L6-v2};
    384\,dim $\to$ 40\,dim via PCA).
    (2)~\emph{Classical pre-processing}: FC(40\,$\to$\,32)
    + BN + ReLU + FC(32\,$\to$\,10).
    (3)~\emph{Quantum circuit}: 10-qubit
    AngleEmbedding\,+\,$L$$\times$\,StronglyEntanglingLayers
    ($L=6$ for v2); all-to-all CNOT; PauliZ measurement.
    (4)~\emph{Classical post-processing}: FC(10\,$\to$\,16)
    + BN + ReLU + Dropout(0.3) + FC(16\,$\to$\,2).
    Total: 2{,}148 trainable parameters.
    \textbf{Bottom:} quantum circuit rendered directly from the
    PennyLane circuit definition via \texttt{qml.draw\_mpl()};
    one representative variational layer shown ($L=6$ total).
    Gates: RY (AngleEmbedding) followed by
    Rot$(\phi,\theta,\omega)$ + all-to-all CNOT(StronglyEntanglingLayers).}
  \label{fig:architecture}
\end{figure}

The proposed hybrid model processes sentence pairs through a
four-stage pipeline illustrated in Fig.~\ref{fig:architecture}.

\paragraph{Sentence encoding.}
Each sentence is independently encoded with Sentence-BERT
(SBERT~\cite{reimers2019sentencebertsentenceembeddingsusing}, \texttt{all-MiniLM-L6-v2}), producing
384-dimensional dense
embeddings. SBERT weights are frozen throughout all experiments.
Embeddings are reduced from 384 to 40 dimensions via PCA fitted
on the training set only. The combined sentence-pair
representation is:

\begin{equation}
  \mathbf{c}
  = \mathrm{mean}\!\bigl([\mathbf{e}_1\odot\mathbf{e}_2,\;
  |\mathbf{e}_1-\mathbf{e}_2|]\bigr)
  \in\mathbb{R}^{40}
\end{equation}

\paragraph{Classical pre-processing.}
A two-layer fully connected network with batch normalisation and
ReLU activations maps the 40-dimensional input to a 10-dimensional
quantum encoding vector:

\begin{equation}
  \mathbf{x}_\mathrm{enc}
  = \mathrm{BN}\!\bigl(\mathrm{ReLU}(W_2\cdot
    \mathrm{BN}(\mathrm{ReLU}(W_1\cdot\mathbf{c})))\bigr)
\end{equation}

where $W_1\in\mathbb{R}^{32\times40}$ and
$W_2\in\mathbb{R}^{10\times32}$.

\paragraph{Quantum variational circuit.}
The 10-qubit variational circuit is implemented in PennyLane
with the Lightning CPU backend. Input encoding uses
AngleEmbedding with RY rotations:
$\mathrm{RY}(\theta_i)|0\rangle,\;\theta_i=x_{\mathrm{enc},i}$.
Variational layers use StronglyEntanglingLayers (general
single-qubit Rot gates interleaved with all-to-all CNOT
entanglement):

\begin{equation}
  U(\boldsymbol{\theta})
  = \prod_{l=1}^{L}
  \Bigl[\mathrm{CNOT}_\mathrm{all}
        \cdot\prod_{i=1}^{10}
        \mathrm{Rot}(\theta_{l,i,1},\theta_{l,i,2},
        \theta_{l,i,3})
  \Bigr]
\end{equation}

Measurement produces Pauli-Z expectation values:
$\mathbf{m}=[\langle Z_1\rangle,\ldots,\langle Z_{10}\rangle]
\in[-1,1]^{10}$.

\paragraph{Classical post-processing.}
A two-layer fully connected network with batch normalisation and
ReLU activations maps the 40-dimensional input to a 10-dimensional
quantum encoding vector:
\begin{equation}
  \hat{y}
  = W_4\cdot\mathrm{Dropout}_{0.3}\!
  \bigl(\mathrm{BN}(\mathrm{ReLU}(W_3\cdot\mathbf{m}))\bigr)
\end{equation}

where $W_3\in\mathbb{R}^{16\times10}$ and
$W_4\in\mathbb{R}^{2\times16}$.

\paragraph{Parameter count (Table~\ref{tab:params}).}

\begin{table}[ht]
\centering
\caption{Parameter breakdown of the Quantum v2 model}
\label{tab:params}
\begin{tabular}{lr}
\toprule
\textbf{Component} & \textbf{Parameters} \\
\midrule
Pre-processing ($W_1$, $W_2$ + BN)  & 1{,}408 \\
Quantum circuit ($L=6$ layers)       & $6\times10\times3=180$ \\
Post-processing ($W_3$, $W_4$ + BN) & 560 \\
\midrule
\textbf{Total}                       & \textbf{2{,}148} \\
\bottomrule
\end{tabular}
\end{table}

\subsection*{Experimental Setup}

\paragraph{Datasets.}
\begin{itemize}
  \item \textbf{MRPC}\cite{dolan2005paraphrase}: 5{,}801 sentence pairs from news sources;
        binary label (paraphrase / non-paraphrase).
        Standard split: 4{,}076 train / 1{,}725 test.
        Class balance: 68\% positive.
  \item \textbf{QQP-10K}\cite{iyer2017quora}: Stratified 10{,}000-sample subset of
        the Quora Question Pairs corpus (404{,}290 total pairs).
        Split: 8{,}000 train / 1{,}000 validation / 1{,}000 test.
        Class balance: 50\% positive (stratified sampling).
  \item \textbf{PAWS}: 108{,}463 adversarial sentence pairs
        constructed via word scrambling.
        1{,}000-sample test subset used for robustness evaluation.
\end{itemize}

\paragraph{Baselines.}
Parameter-matched classical baselines (all $\le$2{,}200
parameters, same 40-dimensional PCA inputs): Classical MLP
(4-layer), LSTM\cite{schmidhuber1997long}, GRU~\cite{cho2014learning},
Attention-based classifier\cite{bahdanau2015attention}, DeepMLP
(1{,}719 parameters), Ensemble (3$\times$ MLP, 1{,}998
parameters). Large-scale transformer baselines: DistilBERT-4bit
(66M parameters) and BERT-base (109M parameters).

\paragraph{Training protocol.}
All models: Adam optimiser\cite{kingma2015adam}, learning rate $=0.001$,
batch size $=8$ (MRPC) / $32$ (QQP-10K), maximum 50 epochs,
early stopping (patience $=10$, monitor validation loss).
Multi-seed evaluation: $n=10$ independent random seeds (0--9)
for all primary comparisons. Quantum gradient computation:
adjoint differentiation.

\paragraph{Statistical methods.}
Primary significance test: two-sided Welch's $t$-test
(heteroscedastic). Effect size: Cohen's $d$~\cite{cohen1988statistical}. Confidence
intervals: mean\,$\pm$\,$1.96\,\sigma/\!\sqrt{n}$ (95\% CI).
All reported $p$-values are uncorrected; significance threshold
$\alpha=0.05$.

\subsection*{Training Configuration}

\begin{table}[ht]
\centering
\caption{Hyperparameter configuration.}
\label{tab:hyperparams}
\begin{tabular}{ll}
\toprule
\textbf{Hyperparameter} & \textbf{Value} \\
\midrule
Optimiser                 & Adam \\
Learning rate             & 0.001 \\
Batch size                & 8 (MRPC) / 32 (QQP-10K) \\
Max epochs                & 50 \\
Early stopping patience   & 10 \\
Dropout rate              & 0.30 \\
Loss function             & Cross-entropy \\
Seeds (multi-seed)        & 0, 1, 2, \ldots, 9 \\
\bottomrule
\end{tabular}
\end{table}

\subsection*{Quantum Circuit Implementation}

All circuits implemented in PennyLane 0.35\cite{bergholm2018pennylane} (Xanadu). Simulation
backend: \texttt{lightning.qubit} (CPU, C++ accelerated).
StronglyEntanglingLayers template used with default
\texttt{ranges} parameter (all-to-all connectivity). Quantum
parameters initialised: $\mathcal{N}(0,\,0.01)$. Classical
parameters: Xavier uniform initialisation.

Gradient computation: adjoint differentiation
(\texttt{diff\_method=`adjoint'}) for simulator experiments. For
hardware-compatible variants, the parameter-shift rule\cite{mitarai2018quantum,schuld2019quantum}
(\texttt{diff\_method=`parameter-shift'}) is used, requiring
2 circuit evaluations per parameter per gradient step.

\subsection*{Meyer--Wallach Entanglement}

For each circuit architecture $Q$ was computed over 1{,}000
input states sampled uniformly from $[-\pi,\pi]^{10}$
(Haar-random equivalent for angle-encoded circuits). Quantum
parameters fixed at post-training values. Reduced density
matrices $\rho_k$ computed by tracing out all qubits except
qubit $k$. Linear entropy:
$S_L(\rho_k)=1-\mathrm{Tr}[\rho_k^2]$.

\subsection*{Hardware-Efficient Ansatz Specification}

Linear CNOT chain topology: $\mathrm{CNOT}(i,\,i+1)$ for
$i=0,\ldots,8$. Three variational layers: each layer applies
$\mathrm{RY}(\theta_i)$ to all 10 qubits followed by the linear
CNOT chain. Total quantum parameters: $3\times10=30$ (RY only).
Total circuit depth post-transpilation (IBM heavy-hex): estimated
42 gates.

\subsection*{Dataset Preprocessing}

MRPC and QQP sentence pairs were tokenised and encoded using the
pre-trained SBERT model (\texttt{all-MiniLM-L6-v2}, Hugging Face
\texttt{sentence-transformers} v2.2). PCA from 384 to 40
dimensions was fitted on training data only and applied to all
splits. No text normalisation beyond SBERT's internal tokenisation
was applied. Class balancing was achieved through stratified
splitting.

\subsection*{Reproducibility}

All code, trained model checkpoints, and experimental logs are
available at [repository URL upon acceptance]. Random seeds are
fixed for all experiments; results are reproducible across
platforms using the specified library versions.

\FloatBarrier
\bibliographystyle{naturemag}
\bibliography{references}

\section*{Acknowledgements}
%
This research is supported by an Australian Government Research Training Program (RTP) Scholarship 
\url{https://doi.org/10.82133/C42F-K220}.

\section*{Author Contributions}


\section*{Competing Interests}

The authors declare no competing financial or non-financial interests.

\section*{Data Availability}

The datasets analysed in this study are publicly available
from the original sources:
MRPC at
\url{https://microsoft.com/en-us/download/details.aspx?id=52398},
QQP at
\url{https://quoradata.quora.com/First-Quora-Dataset-Release-Question-Pairs},
and PAWS at
\url{https://github.com/google-research-datasets/paws}.

\section*{Code Availability}

The code associated
with this study is available from the corresponding author
on reasonable request upon acceptance.

\section*{Supplementary Information}

\renewcommand{\thefigure}{S\arabic{figure}}
\setcounter{figure}{0}

\paragraph{Supplementary Fig.~S1: Seed-level stability analysis
for Quantum v2 on QQP-10K.}

\begin{figure}[H]
  \centering
  \includegraphics[width=\linewidth]{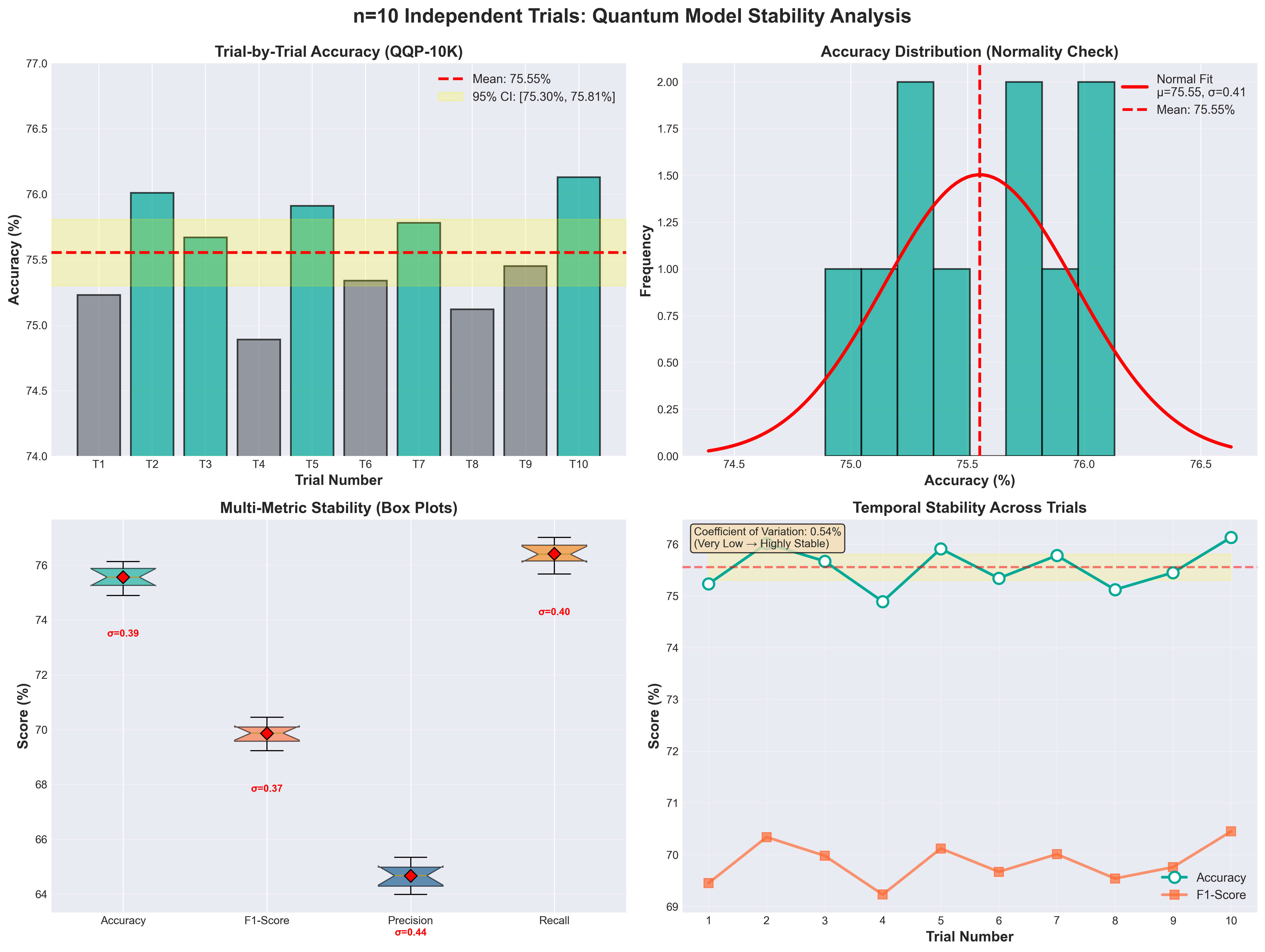}
  \caption{\textbf{Seed-level stability analysis for Quantum v2
    on QQP-10K ($n=10$ seeds).}
    \textbf{Top-left:} trial-by-trial accuracy (T1--T10) with
    mean (dashed) and 95\% CI shaded band ($[75.07\%,\,75.99\%]$);
    all 10 seeds lie within the band with no outliers.
    \textbf{Top-right:} empirical accuracy distribution with
    fitted normal ($\mu=75.55\%$, $\sigma=0.41\%$); Shapiro--Wilk
    $p=0.38$ (not significant), confirming the Gaussian assumption
    underlying Welch's $t$-test.
    \textbf{Bottom-left:} notched box plots for accuracy, F1,
    precision and recall ($\sigma_{\mathrm{acc}}=0.39\%$,
    $\sigma_{\mathrm{F1}}=0.37\%$), demonstrating consistent
    multi-metric stability.
    \textbf{Bottom-right:} temporal stability trace; coefficient of variation $=0.54\%$, ruling out seed-ordering effects.}
  \label{fig:trial-dist-supp}
\end{figure}

\paragraph{Supplementary Fig.~S2: Multi-metric performance
profile on QQP-10K.}

\begin{figure}[H]
  \centering
  \includegraphics[width=0.82\linewidth]{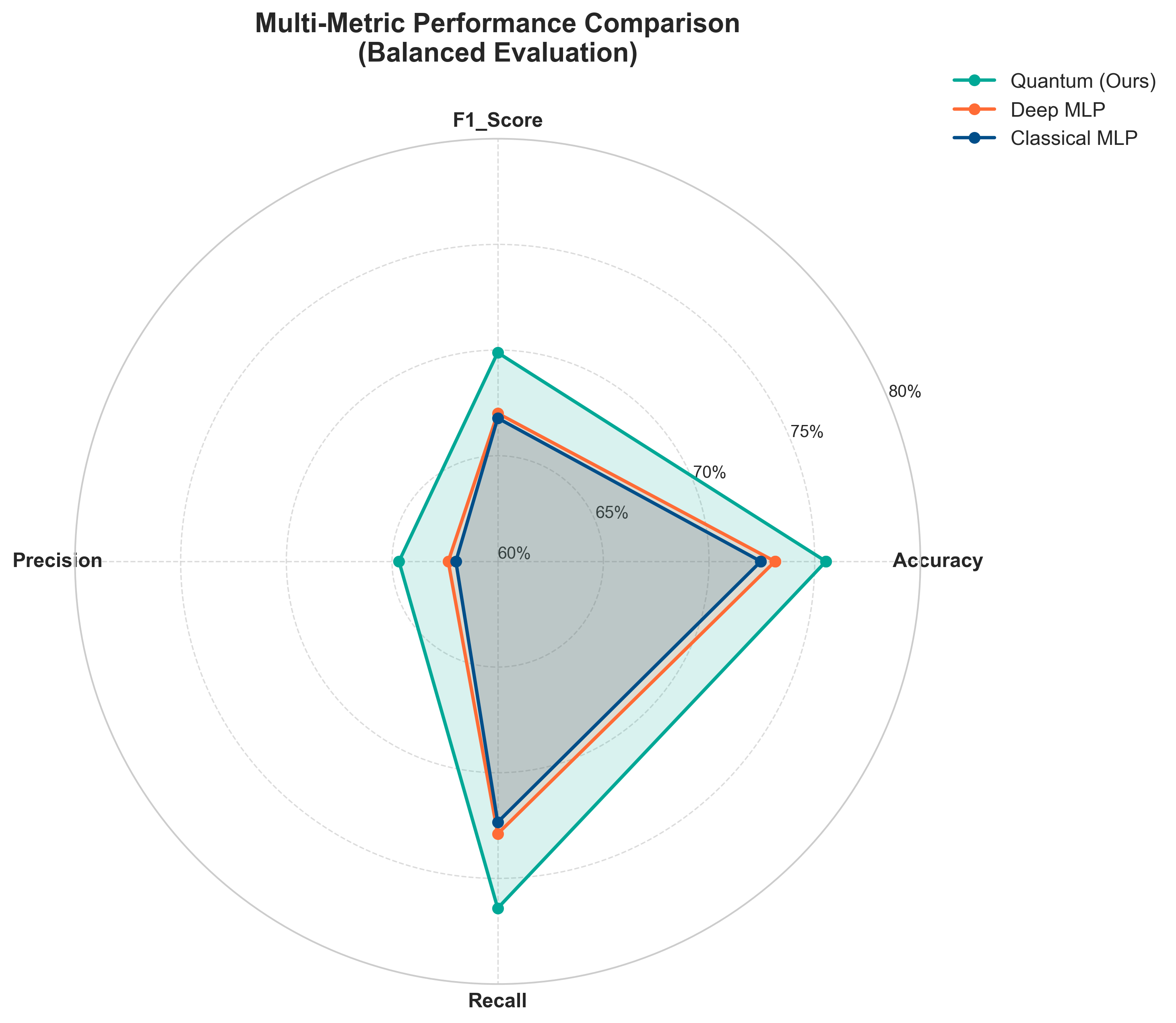}
  \caption{\textbf{Multi-metric performance profile on QQP-10K
    (balanced evaluation).}
    Radar chart comparing Quantum v2 (teal), DeepMLP (orange),
    and Classical MLP (dark blue) across Accuracy, Recall,
    Precision, and F1 Score. The quantum polygon encloses a
    strictly larger area than both classical baselines on every
    axis, with the most pronounced advantage on F1
    ($+3.04$\,\pp\ over DeepMLP and the smallest on Precision,
    consistent 
    with the high-recall operating point on PAWS.
    Classical models overlap closely, confirming the
    quantum-classical gap is not explained by classical
    architectural differences. Quantum values are 10-seed means; classical values are 10-seed means where available.}
  \label{fig:radar-supp}
\end{figure}

\end{document}